\documentclass[a4paper,aip,apl,amsmath,amssymb,reprint,superscriptaddress]{revtex4-1}
\usepackage{graphicx}
\usepackage{amsmath}
\usepackage{dcolumn}
\usepackage{bm}
\usepackage{bbold}
\usepackage{color}
\usepackage{amsfonts}
\usepackage{amssymb}
\usepackage{mathrsfs}
\usepackage{tabularx}
\usepackage{braket}
\usepackage{mathtools}
\usepackage{soul}			

\usepackage{caption}
\usepackage{subcaption}

\begin{document} 
\title{Strong Coupling of 3D Cavity Photons to Travelling Magnons At Low Temperatures}
\author{Maxim Goryachev}
\affiliation{ARC Centre of Excellence for Engineered Quantum Systems, School of Physics, University of Western Australia, 35 Stirling Highway, Crawley WA 6009, Australia}

\author{Mikhail Kostylev}
\affiliation{Magnetisation Dynamics and Spintronics Group, School of Physics, University of Western Australia, 35 Stirling Highway, Crawley WA 6009, Australia}

\author{Michael E. Tobar}
\email{michael.tobar@uwa.edu.au}
\affiliation{ARC Centre of Excellence for Engineered Quantum Systems, School of Physics, University of Western Australia, 35 Stirling Highway, Crawley WA 6009, Australia}

\date{\today}


\begin{abstract}

We demonstrate strong coupling between travelling magnons in an Yttrium Iron Garnet film and 3D microwave cavity photons at milli-Kelvin temperatures. The coupling strength of $350$MHz or $7.3$\% of resonance frequency is observed. The magnonic subsystem is represented by the Damon-Eshbach magnetostatic surface wave with a distribution of wave numbers giving the linewidth of 15MHz. The ways to improve this parameter are discussed. The energy gap in the spectrum given by the Zeeman energy and the shape-anisotropy energy in the film geometry give rise to a significant asymmetry of the double peak structure of the photon-magnon avoided level crossing.
A structure of two parallel YIG films is investigated using the same re-entrant magnetostatic surface wave transducer revealing a higher order magnon modes existing in both films. Combination of a multi-post re-entrant cavity and multiple films is a potential base for engineering both magnon and photon spectra. 

\end{abstract}

\maketitle

In recent years, ferromagnetic materials in cryogenic conditions have become a subject of intensive study with various potential applications. Firstly, magnon modes in highly regular ferrimagnetic crystals such as Yttrium Iron Garnet (YIG) has been widely considered as a candidate for matter part in Quantum Electrodynamics (QED) experiments and its applications\cite{Soykal:2010ly,PhysRevLett.113.083603,Goryachev:2014ab,Tabuchi:2014aa,Morris:2017aa} as well as some related subjects such as optomechanics\cite{Zhang:2016aa,Liu:2016aa}. These proposals are motivated by unique properties of magnons such as high regularity of the crystal and high spin density in this material resulting in very narrow magnon and photon linewidths and extremely strong coupling strengths to photons. Secondly, in addition to applications in quantum science, YIG is considered as a detector material for spin-coupled dark matter searches, particularly galactic axions\cite{sikivie}, where large number of spins and low decoherence increase the overall detector sensitivity\cite{Barbieri:2017aa}. In addition to that, low photon losses, strong nonlinear effects and broken time reversal symmetry may result in a number of other interesting applications in low temperature and condensed matter physics. 

Dictated by the widespread room temperature applications of YIG in microwave tunable filters and oscillators, the most common form of this material is a sphere. This form together with the regularity of the crystals themselves guarantee not only narrow linewidths but also suppression of higher order magnon modes. These modes are typically unwanted effects suppressing the overall system performance. Indeed, imperfect crystals exhibit asymmetries that result in coupling of uniform magnetic fields to nonuniform modes. At room temperature, the problem is solved by improving symmetry of the crystals with state-of-the-art manufacturing technologies.  At cryogenic temperatures these techniques do not work, as cooling induces significant strains and stresses resulting in additional asymmetries that cannot be taken into account during manufacturing. For this reason, all cryogenic tests of YIG spheres demonstrate significant coupling to spurious modes\cite{Goryachev:2014ab,Tabuchi:2015aa,Wang:2016aa,Lambert:2015aa,Zhang:2014aa,Zhang:2017aa} often within an avoided level crossing with the uniform magnon resonance. In real applications, this effect will decrease overall coherence through coupling to extra degrees of freedom.

To solve the problem, one might consider other geometrical forms of YIG crystals. Particularly, rectangular blocks and spheroids with flat surfaces have been considered to decrease the degeneracy of the modes. These attempts did not result in significant improvement in the magnon mode structure. Additionally the associated fabrication techniques are based on hand polishing of YIG spheres or immature crystal cutting. In this work, we consider epitaxial YIG films, a well established crystal growth technology, as a building block for spectrally pure magnonic parts of QED systems and associated perspectives in applications of this material in low temperature physics. Utilisation of YIG material in the form of films is also motivated by the fact that it has been employed as a platform to study nonlinear effects in magnetic systems\cite{Wu:2005aa,1983JETPL}, magneto-optical phenomena\cite{Wettling:1973aa,Anshakov:1990aa,Tsai:1985aa,Matyushev:1995aa}, light/microwave interaction\cite{Klingler:2016aa}, Bose-Einstein condensates of magnons\cite{Demokritov:2006aa}, transport phenomena\cite{Cornelissen:2015aa} as well as magnon-photon coupling properties of YIG films have been studied in microwave regime at room temperature\cite{Bhoi:2014aa,Stenning:2013aa,Zhang:2016ab}. Moreover, it has been proposed theoretically that magnetic films in 3D cavities and gratings may result in some interesting effects\cite{Dodonov:2017aa,Maksymov:2014aa}.

\begin{figure}[h!]
     \begin{center}
            \includegraphics[width=0.35\textwidth]{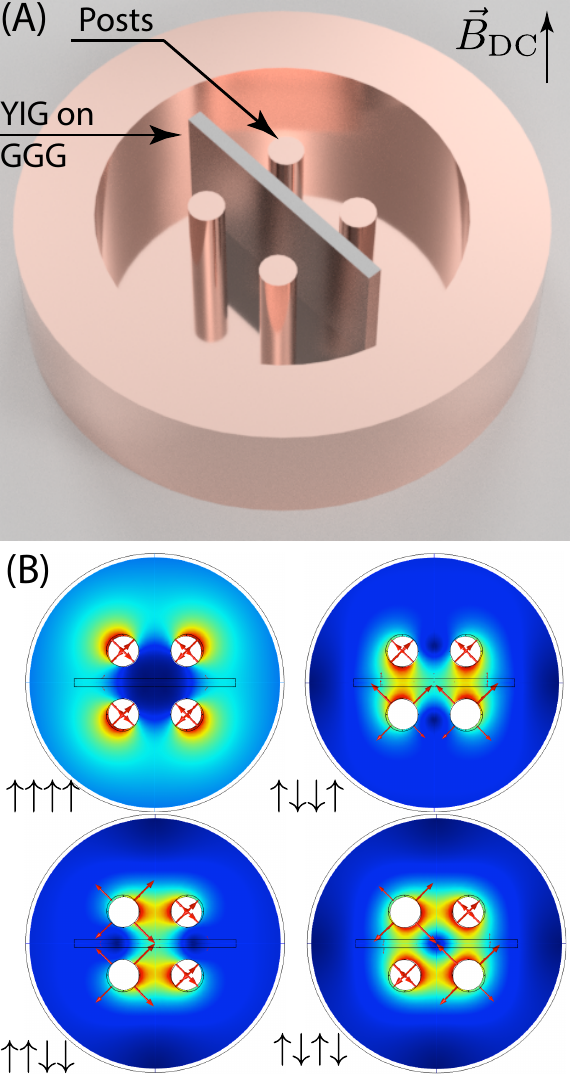}
            \end{center}
    \caption{(A) Four post re-entrant cavity with an YIG film deposited on a GGG crystal. (B) Magnetic field in the cavity plane for four re-entrant modes. }%
   \label{cavity}
\end{figure}

To demonstrate the possibilities of YIG film, we employ a multi-post re-entrant cavity\cite{Goryachev:2015aa} that allows precise spectrum and magnetic field pattern engineering\cite{Goryachev:2014ab,Goryachev:2016aa}. The number of posts in this work is limited to four, and the cavity is oriented in such a way that external DC magnetic field is applied along the direction of the posts, such that the resonant microwave magnetic field is normal for all the cavity re-entrant modes (see Fig.~\ref{cavity} (A)). The cavity is 5mm heigh and 15mm in diameter with 4 mm distance between the posts in both directions. The YIG film is positioned between two pairs of the microwave posts along them and the external DC field. Thus, the microwave $B$ field components could be either normal or parallel to the plane depending on the mode structure. Four re-entrant post result in four microwave modes in the cavity that can be classified by the direction of currents through the posts at the same instance of time. Firstly, the mode with the uniform current distribution ($\uparrow\uparrow\uparrow\uparrow$) corresponds to the lowest frequency resonance. This mode expels the magnetic field outside of the space between the posts as shown in Fig.~\ref{cavity} (B). Secondly, two dipole type modes with pair-wise distribution of currents have to be degenerate in frequency in a perfectly symmetric (under 90 degrees rotation) cavity but exhibit large frequency splitting in presence of large disturbance such as the YIG crystal. The first dipole mode ($\uparrow\downarrow\downarrow\uparrow$) concentrates the magnetic field in the YIG crystal as the currents in the posts on both sides of the film point in opposite directions. The situation is reversed for the second dipole mode ($\uparrow\uparrow\downarrow\downarrow$) where oppositely oriented currents are on the same side of the film. 
Thirdly, the quadrupole mode ($\uparrow\downarrow\uparrow\downarrow$) is characterised by alternating directions of displacement currents under each posts along both directions in the cavity plane giving the highest frequency mode and some magnetic field concentrated in the film.

To observe photon-magnon interaction in the described above system at low temperatures, the four post cavity made of Oxygen Free Copper was attached to the $20$mK stage of a dilution refrigerator inside a superconducting magnet. The sample used in the experiment is (111) YIG crystal of $4\mu$m thickness deposited on a GGG square substrate with $10$mm side and 0.5mm overall thickness. The crystal inserted in a slot such that only $5$mm of its height is situated inside the cavity. Thus, the total volume of the YIG material inside the cavity is $2\times 10^{-10}$ m$^3$ that is more than order of magnitude less than a sphere of radius $1$mm. With the given dimensions, the distance between the surface of the posts and the crystals is on the order 0.75mm.

\begin{figure*}[hbt!]
     \begin{center}
            \includegraphics[width=1\textwidth]{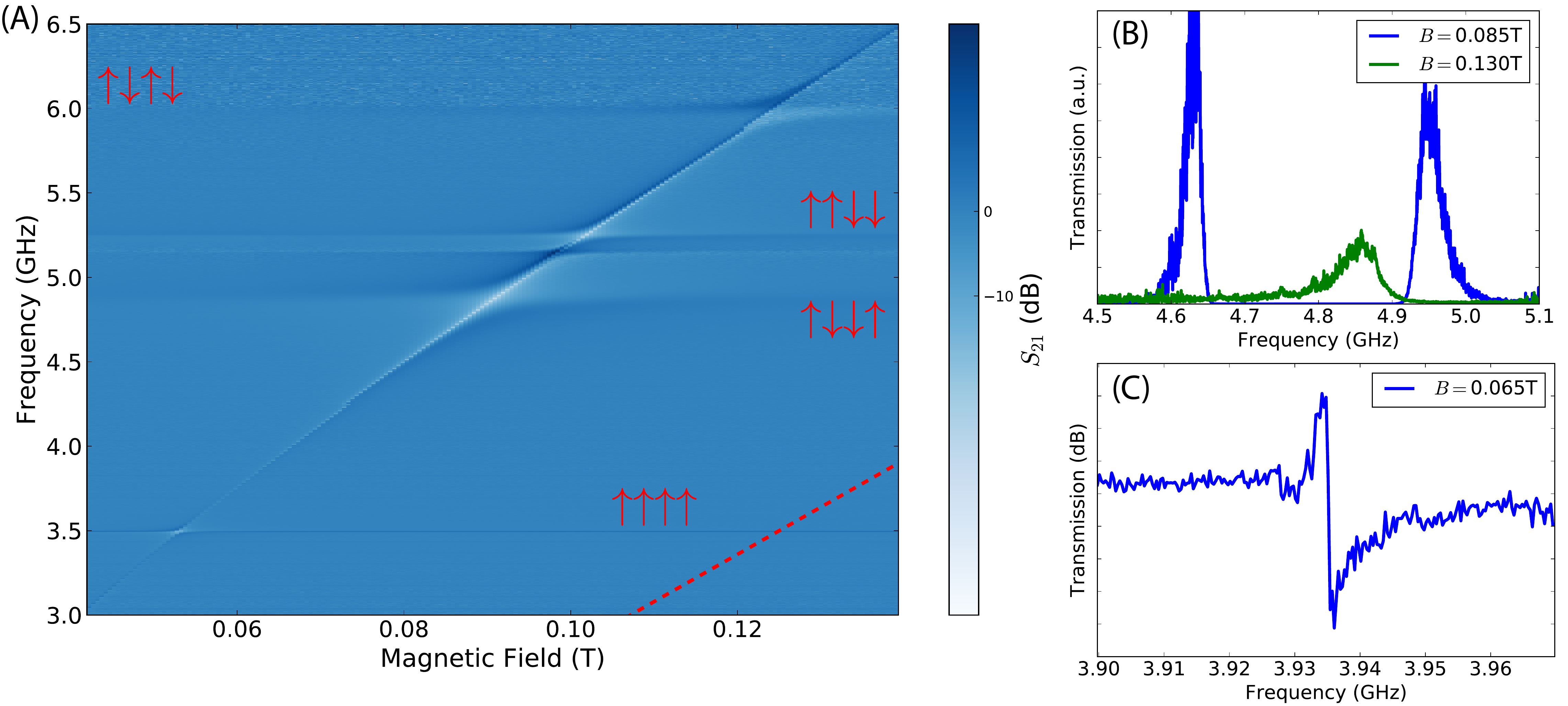}
            \end{center}
    \caption{(A) Transmission through the four post cavity with the YIG/GGG crystal as a function of external magnetic field. (B) Mode splitting between the $\uparrow\downarrow\downarrow\uparrow$ cavity resonance and the YIG film mode. (C) Magnon line shape outside of the observed avoided level crossings.}%
   \label{firstresult}
\end{figure*}

The system is characterised using the transmission method with the incident signal cryogenically attenuated by 40dB at 4K and 20mK stages. The transmitted signal is amplified by a low noise cryogenic and room temperature amplifiers. The transmission through the system is measured as a function of the DC magnetic field and is shown in Fig~\ref{firstresult} (A). The spectrum demonstrates four cavity modes described above together with a single YIG film magnon that is tuned over the whole frequency range with the external DC magnetic field. For reference, the dashed line in Fig~\ref{firstresult} (A) demonstrates the tuning dependence of a spin system with Land\'{e} g factor of 2 (or gyromagnetic ratio $\frac{\gamma}{2\pi}$ of approximately $28$GHz/T) and vanishing zero field splitting. The observed magnon line fits the Kittel formula\cite{SW1}:
\begin{equation}
\begin{array}{l}
\displaystyle f = \frac{\gamma}{2\pi}\sqrt{B_\text{DC}(B_\text{DC} + \mu_0M_s)},
\label{kittel}
\end{array}
\end{equation}
of an in-plane magnetised film where the saturation magnetisation $\mu_0M_s = 0.236$T and gyromagnetic ratio is $28.3$GHz/T.

When resonance frequencies of cavity modes and magnon resonance interact, the system exhibits avoided level crossing. The corresponding mode splitting for this interaction is shown in Fig.~\ref{firstresult} (B). The strongest interaction between subsystems is observed for the dipole resonance antisymmetric around the film and estimated to be on the order of $350$MHz or $7.3$\%. This value is less than that observed for the field focusing re-entrant cavities and YIG spheres\cite{Goryachev:2014ab} but exceeds some other sphere experiments\cite{Zhang:2017aa,Tabuchi:2014aa,Tabuchi:2015aa,Zhang:2014aa}. 

The magnon mode frequency response outside of any avoided level crossing is shown in Fig.~\ref{firstresult} (C). Based on this response the magnon linewidth may be estimated to be on the order of 15-20MHz. The observed linewidth is inferior to that for state-of-the-art YIG spheres at mK temperatures where linewidths are smaller than 1MHz\cite{Goryachev:2014ab,Tabuchi:2014aa} but usually measured values above 1MHz\cite{Tabuchi:2015aa}. It is on the same order as that for highly doped paramagnetic crystals, which however typically have smaller spin density, larger volumes and smaller couplings. Understanding of the increased linewidth for the magnon resonance for the film may be provided based on the theory in Bhoi et al\cite{Bhoi:2014aa}. In fact, the resonance linewidth for high-quality epitaxial YIG films is just slightly worse than for YIG spheres - usually on the order of $0.5$Oe or $1.4$MHz for the observed gyromagnetic ratio.

In order to observe the intrinsic FMR linewidth for a film, one needs to localise magnetisation dynamics within an area with in-plane sizes significantly smaller than the free-propagation path for travelling spin waves in the film, thus creating conditions for a well-resolved discrete spectrum of standing wave oscillations across the area of localisation of the dynamics. This can be achieved, for instance, by using a small rectangular or circular piece of the film (typically 1 to $2$mm in in-plane size)\cite{Gieniusz:1993aa}. 
Otherwise spin waves excited by a localised microwave source, such as the four posts in the present work or a planar split-ring resonator, as in Bhoi et al\cite{Bhoi:2014aa}, are not bound by the area of excitation and escape the area as traveling waves carrying energy away from the area. Eventually, such magnons decay due to losses in the film within several free-propagation path of spin waves $l_f$. For a 4$\mu$m YIG film, such as one used in the present experiment its value is $l_f=3$mm.

The magnon waves excited in the present experiment are plane waves that propagate in the film in both directions perpendicular to the posts. As the static magnetic field is applied along the posts, the type of wave which is excited in this configuration is the Damon-Eshbach (DE) Magnetostatic Surface Wave (MSSW)\cite{Damon:1961aa}. As for any traveling wave, the dispersion relation, or energy spectrum, is continuous. This continuity gives rise to asymmetry in the double peak structure as observed by the 4-post MSSW transducer and depicted in Fig.~\ref{firstresult}(B): their lower-frequency slopes are significantly steeper than the higher-frequency ones. The explanation of this phenomenon lies in their different physical origins\cite{Bhoi:2014aa}. The steeper slope is formed because of the energy gap in the spectrum given by the Zeeman energy and the shape-anisotropy energy in the film geometry. There are no travelling waves excitations in the energy gap, therefore the microwave power absorbed by the system drops abruptly at the edge of the gap giving rise to this steep slope. 
The edge of the gap corresponds to the zero spin-wave wave number $k$ and its frequency $f_g$ is given in Eq.~(\ref{kittel}). With an increase in the frequency $f$, $k$ increases, following the DE dispersion law. With an increase in $k$, the efficiency of excitation of MSSW by the microwave current in the posts gradually drops forming the more gradual higher-frequency slopes of the peaks in Fig.\ref{firstresult}(B), the maximum excited wavenumber $k_\text{max}$ being dependent on the post diameter and their distance to the film surface.
Another important feature of the observed interaction is a week oscillatory pattern at the top of this slope. The origins of this pattern is due to the presence of two pairs of the posts which are equivalent to two opposite sides of the square split-ring\cite{Bhoi:2014aa}. These sides independently excite pairs of partial spin waves in two opposite directions. Because of coherence of the partial waves, the waves interfere in the film space between the posts which leads to the oscillatory interference pattern seen on top of the slopes.  

The observed phenomena suggest possible ways to decrease the peak widths and, hence, to increase the magnon photon coupling. Firstly, one may use a thinner film. Indeed, for small wave numbers the MSSW frequency scales as $df=f-f_g\sim kt$, where $t$ is the film thickness. Form this relation, it follows that for the same $k_\text{max}$, the band of excited waves $df$ becomes smaller due to a decrease in $t$. Unfortunately, the decrease in $t$ also decreases the volume of the magnetic material in the cavity, hence, the coupling to the cavity microwave field. 
One possible way to get around this problem is by placing multiple (highly identical) films parallel to each other between the posts, separated by a distance on the order of $1$~mm from each other in order to ensure that the film do not couple by their dipole fields. 
Secondly, one can can pattern the film into an array of dipole uncoupled squares with edge sizes on the order of $1$~mm\cite{Gieniusz:1993aa}.  

\begin{figure}[hbt!]
     \begin{center}
            \includegraphics[width=0.5\textwidth]{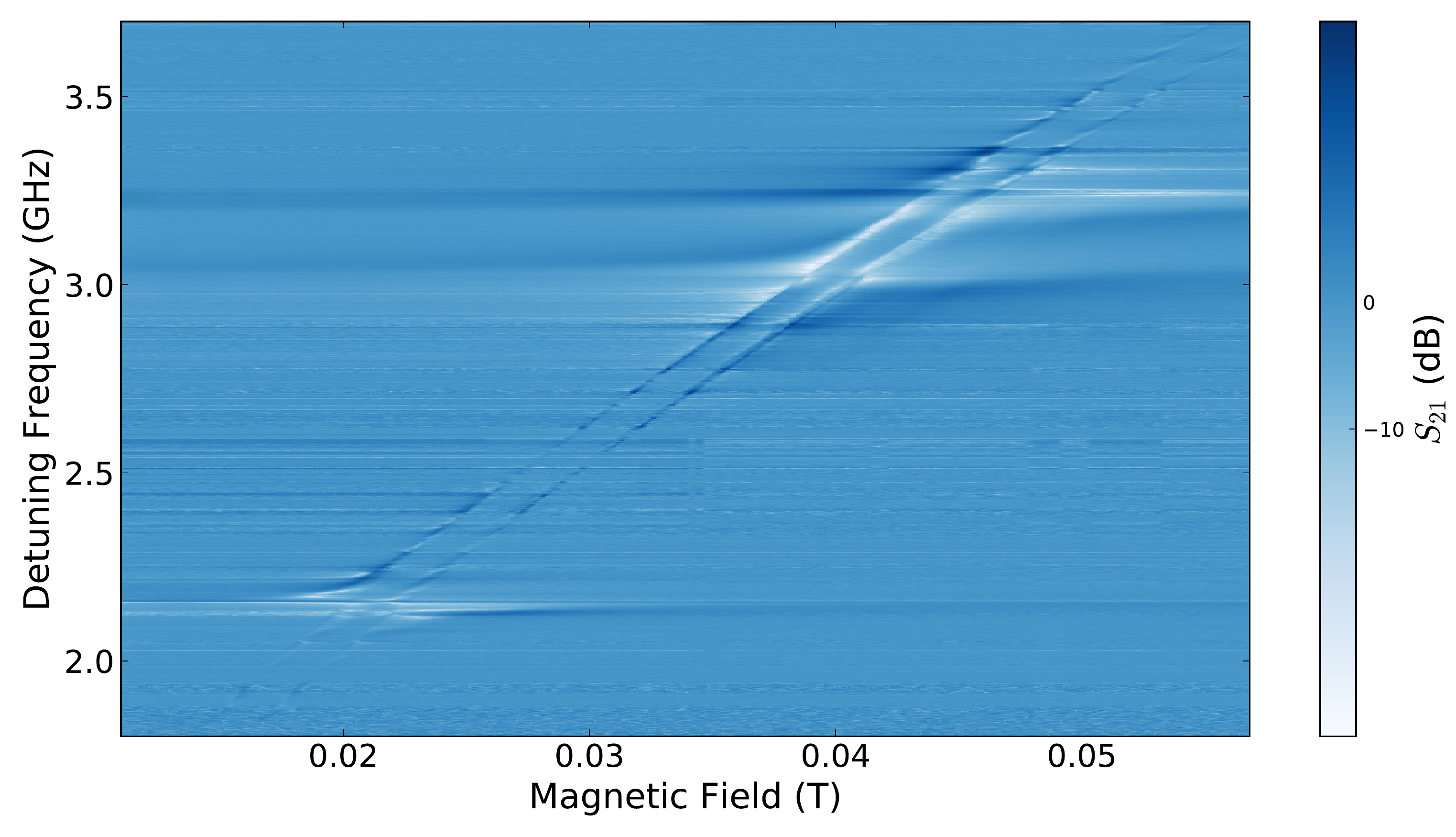}
            \end{center}
    \caption{Transmission through the four post cavity with the two YIG/GGG crystals stacked together as a function of external magnetic field.}%
   \label{doublefilm}
\end{figure}

To demonstrate versatility of the 3D cavity coupled to an YIG film, we investigate a system with two parallel films in the same cavity. Both films are nominally identical and deposited on different GGG substrates of the same size. The crystals are stuck together in a firm mechanical contact and put in the same slot of the same re-entrant cavity. The frequency response of this system is shown in Fig.~\ref{doublefilm} where two distinct parallel magnon resonance lines are visible. Fitting these lines to the same the Kittel formula reveals that parameters for the right resonance are exactly the same as for the single film case, there as the best fit for the left resonance gives $\mu_0M_s = 0.36$T and gyromagnetic ratio of $24.5$GHz/T and considerable deviation form the data trace at lower magnetic fields. Such significant difference in parameters cannot be accounted for the nonuniformity of the film that is known to have a larger volume of the uniform field or discrepancies in the film parameters. Thus, it suggests that the left resonance manifests higher order magnon waves existing in both films.
Such solution with multiple films allows to engineer not only photon cavity spectra as demonstrated with re-entrant cavities\cite{Goryachev:2014ab,Goryachev:2016aa}, but also magnon frequency distributions. For instance, this feature may be used to lower the required magnetic field in the applications requiring superconducting parts\cite{Tabuchi:2014aa,Morris:2017aa}. Additionally such hybrid structures might be used for applications where a number of spins is of primary importance such as dark matter detection \cite{sikivie,Barbieri:2017aa}.

In conclusion, we demonstrated the strong coupling regime between 3D cavity photons and YIG film magnons at 20mK. The demonstrated coupling exceeds that of many YIG sphere experiments with a room for further improvement. Although, the magnon linewidth in a film is larger than in a typical sphere, it is less than that for highly doped paramagnetic system with an additional advantage of much stronger coupling. Also, it is demonstrated that by employing multiple films, one can make a more complicated magnon frequency response.

This work was supported by the Australian Research Council grant number CE110001013.

\section*{References}

%

\end{document}